\documentclass[twocolumn,showpacs,preprintnumbers,amsmath,amssymb]{revtex4}

\usepackage{graphicx}% Include figure files
\usepackage{dcolumn}% Align table columns on decimal point
\usepackage{bm}% bold math

\begin{document}

\preprint{APS/123-QED}

\title{Control of many electron states in semiconductor quantum dots by non-Abelian vector potentials }
\author{S.-R. Eric Yang\footnote{ eyang@venus.korea.ac.kr}}
\affiliation{Physics Department, Korea  University, Seoul Korea} 

%\date{\today}

\begin{abstract}
Adiabatic time evolution of degenerate eigenstates of a quantum system provides a means
for controlling  electronic states  since mixing between degenerate levels generates
a matrix Berry  phase. 
In the presence of spin-orbit coupling in n-type semiconductor quantum dots 
the electron Hamiltonian is invariant under time reversal operation
and the many body groundstate  may be doubly degenerate.
This double degeneracy can generate  non-Abelian vector potentials  when odd number of electrons are present.  
We find that the  antisymmetry of many electron wavefunction has no effect on the matrix Berry phase.
We have derived equations that allow one to 
investigate the effect of  electron correlations  by expressing the non-Abelian vector potentials 
for many electron system in terms of single electron non-Abelian vector potentials.

\end{abstract}
\maketitle

\section{Introduction}

Single electron control in semiconductor quantum dots  would be valuable for 
spintronics, quantum information, and spin qubits\cite{Aws,ras0}.
Adiabatic time evolution of degenerate eigenstates of a quantum system provides a means
for controlling individual electrons  since mixing between degenerate levels generates
a matrix Berry  phase\cite{Sha,Wil}. 
It has been shown that universal quantum computation is possible by means of non-Abelian unitary
operations\cite{Zan,Pach}.
Recently several proposals on how to generate matrix Berry  phases have been made in semiconductor 
systems\cite{Bern,Sere2,Sol2,par,yang1}.
In one of these proposals it has been suggested that 
a {\it single electron} in a  n-type semiconductor quantum dot with spin-orbit 
terms  can have matrix Berry phases\cite{yang1}.
All the   discrete energy levels of  quantum dots possess  a double degeneracy 
because  the Rashba and/or Dresselhaus spin orbit coupling terms have time reversal symmetry\cite{val}.
In Ref.\cite{yang1}  it was suggested that the adiabatic transformation
can be performed {\it electrically} by changing the confinement potentials of the quantum dot. 
It was shown   that matrix Berry phases can be  produced only when  the parity  
symmetry of two-dimensional harmonic potentials is broken.  
A degenerate pair of states also exists in quantum rings\cite{yang2}.  The relevant symmetry of this system 
is the combined operation of  time reversal and large gauge transformations.
However, it is often not easy  to populate a quantum dot or ring with just one  electron and usually there 
are  several electrons in it\cite{kou,lee}.
It is not certain that many body groundstates are  doubly degenerate in the presence of time reversal symmetry.
Moreover, it is unclear how the effects of  exchange and correlation many body physics 
change  the non-Abelian vector potentials.  
These issues are investigated in the present paper.

Before we give a summary of the main points of the present paper 
let us give a  brief introduction   of the basic ideas  behind  matrix Berry phases. 
Consider a degenerate pair of states $|\Psi_1\rangle $   and $|\Psi_2\rangle$, which  may be {\it single particle
or many body states}.
If the system is  in a superposition  state $|\Psi(0)\rangle =c_1(0)|\Psi_1(0)\rangle +c_2(0)|\Psi_2(0)\rangle$ 
at time t=0  an adiabatic  evolution  of the parameters
$\lambda_p$ can  transform this state into another state $|\Psi(t)\rangle =c_1(t)|\Psi_1(t)\rangle +c_2(t)|\Psi_2(t)\rangle$ 
after some time $t$. Here the orthonormal basis states $|\Psi_i(t)\rangle$ are the instantaneous eigenstates of the Hamiltonian. 
For a cyclic change with the  period  $T$, represented by a closed contour $C$ in the parameter space,
the states $|\Psi_1(T)\rangle$ and  $|\Psi_2(T)\rangle$ return to the initial states $|\Psi_1(0)\rangle$ and 
$|\Psi_2(0)\rangle$, but the coefficients
$c_1(T)$ and $c_2(T)$ may not return the initial values.  In such a case a $2\times2$ matrix Berry phase
(non-Abelian Berry phase) $\Phi_C$ is generated
\begin{eqnarray}
\left(
\begin{array}{c}
c_1(T) \\
c_2(T)
\end{array}
\right)=\Phi_C
\left(
\begin{array}{c}
c_1(0) \\
c_2(0)
\end{array}\right).
\end{eqnarray}
The expansion coefficients  $c_1(t)$ and $c_2(t)$   satisfy the time-dependent Schr$\ddot{o}$dinger equation
\begin{eqnarray}
i \hbar \dot{c}_i=-\sum_j A_{i j} c_j \qquad i=1,2.
\label{eq:time_Schrod}
\end{eqnarray}
The matrix elements $A_{ij}$ are given by
$A_{ij}= \hbar \sum_p(A_p)_{i,j}\frac{d\lambda_p}{dt}$, where the
sum over $p$ in $A_{ij}$ is meant to be the sum  over $\lambda_p$.
The  non-Abelian vector potentials
$A_p$ are  $2\times2$ matrices 
\begin{eqnarray}
(A_p)_{i,j}=
i \langle\Psi_i|\frac{\partial\Psi_j}{\partial\lambda_p}\rangle .
\label{eq:vector_pot}
\end{eqnarray}
Note that the wavefunctions $\Psi_i$  and $\Psi_j$ are  {\it degenerate}.
The role of the non-Abelian vector potentials is to keep  the energy degeneracy intact while the system parameters
change.

\begin{figure}[hbt]
\begin{center}
\includegraphics[width = 0.3 \textwidth]{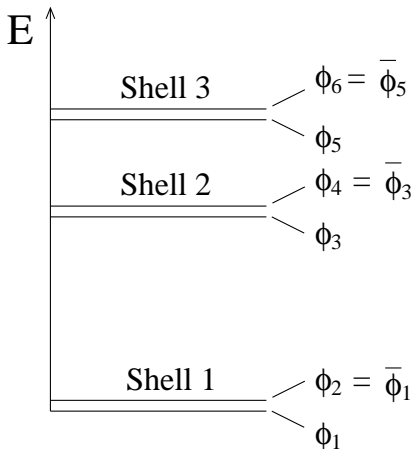}
\caption{Single electron eigenstates form degenerate energy shells.  In each shell electron wavefunctions form a time reversed pair.
}
\label{fig:single}
\end{center}
\end{figure}

Our investigation shows  that when even number of electrons are present the many body groundstate may not be doubly degenerate.
However, when odd number of electrons are present the groundstate may be  doubly degenerate.
In the simplest possible approximation the antisymmetry of many body wavefunctions can be  ignored.
Then the degenerate groundstates for three electrons are given by 
$|\Psi_1\rangle= \phi_1(r_1)\phi_2(r_2)\phi_3(r_3)$ and
$|\overline{\Psi}_1\rangle=\phi_1(r_1)\phi_2(r_2)\phi_4(r_3)$,
where $\phi_i$ are single electron wavefunctions, see Fig.\ref{fig:single}.
In this case  one can show that the solutions of the time dependent Schroedinger equation, Eq.(\ref{eq:time_Schrod}),
 have the form
\begin{eqnarray}
\left(
\begin{array}{c}
c_{1}(t) \\
c_{2}(t)
\end{array}
\right)=
e^{\frac{i}{\hbar}f(t)}\left(
\begin{array}{c}
B_{1}(t) \\
B_{2}(t)
\end{array}
\right).
\label{eq:berry}
\end{eqnarray}
The function $f(t)$  depends on the single electron non-Abelian vector potentials of the {\it first} shell, and    
the functions $B_{1}(t)$ and $B_{2}(t)$ depend on the single electron non-Abelian vector potentials of 
the {\it second} shell,  see Fig.\ref{fig:single}
(The function  $f(t)$ is periodic with the period $T$ ). 
The probability to find an electron in the state
$|\phi_3\rangle (|\phi_4\rangle)$ is $|c_1(T)|^2=|B_1(T)|^2 (|c_2(T)|^2=|B_2(T)|^2 )$.
So these probabilities 
are  {\it independent} of the phase factor $e^{\frac{i}{\hbar}f(t)}$, i.e., of  the non-Abelian vector potentials of the first 
shell.

\begin{figure}[hbt]
\begin{center}
\includegraphics[width = 0.4 \textwidth]{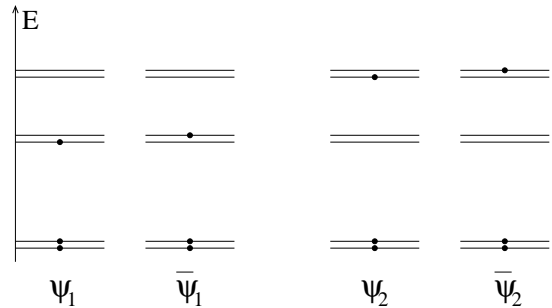}
\caption{Four many body basis states in the three electron Hilbert space. 
The basis pair ($\Psi_1,\overline{\Psi}_1$) or $(\Psi_1,\overline{\Psi}_1)$ is
a time reversed pair and has the same total single electron energy.
So do $\Psi_2$ and $\overline{\Psi}_2$.   Black dots represent electrons.
}
\label{fig:many}
\end{center}
\end{figure}

In the next simplest possible approximation the antisymmetry of many body wavefunctions can be
included.  In this case the doubly degenerate many body groundstates
may be taken to be  two Slater determinant wavefunctions, $\Psi_1$ and $\overline{\Psi}_1$, 
that are time reversed states of each other, see Fig.\ref{fig:many}.
It will be shown below  that  the antisymmetry of these  Slater determinant wavefunctions does {\it not} change the
results obtained above for the matrix Berry phase.
In addition to  the antisymmetry  one can  include correlation effects.   This can be done by including four basis vectors
$\Psi_1$ ,$\overline{\Psi}_1$, $\Psi_2$ and $\overline{\Psi}_2$, see Fig.\ref{fig:many}.  In this case 
the groundstates are given by linear combinations of these four states and are thus correlated.
It can be shown that  
the non-Abelian gauge potentials for a correlated  electron system are related to
non-Abelian gauge potentials for single electrons, provided that the  
definition of  single electron non-Abelian gauge potentials is extended to  
\begin{eqnarray}
(a_q)_{k,p}=
i \langle\phi_k|\frac{\partial\phi_p}{\partial\lambda_q}\rangle 
\end{eqnarray}
when  $\phi_k$ and $\phi_p$  belong to {\it different} single electron energy shells. 
We will call these {\it inter shell non-Abelian gauge potentials}.

In Sec. II we describe the   Hamiltonian of the system  in detail, and in Sec. III we discuss 
why  matrix Berry phases can be  produced only when  the parity  
symmetry of two-dimensional harmonic potentials is broken.  
A computational scheme to include systematically many electron physics 
is explained in Sec.IV.
The absence of a matrix Berry phase for the two-electron case
is shown in Sec. V.  In Sec.VI we show  that 
a matrix Berry phase is present in the three-electron case.
Correlation effects are discussed in VII and conclusions  are given in Sec. VIII.

\section{Model Hamiltonian}

An electric field $E$ is along the z-axis and electrons are confined in a triangular potential $V(z)$.
When the the width of the quantum well along the z-axis is sufficiently {\it small} we may include only the lowest
subband state along  the z-axis. We denote this  wavefunction by $f(z)$.
From the expectation value
$\langle f(z)|k_z^2|f(z)\rangle =0.8(2m^*eE/\hbar^2)^{2/3}$
we estimate the characteristic length scale along the z-axis: $R_z=1/\sqrt{0.8(2m^*eE/\hbar^2)^{2/3}}$.
The Hamiltonian in the absence of the spin orbit coupling is 
\begin{eqnarray}
H_K=-\frac{\hbar^2\nabla^2}{2m^*}+U(x,y)+V(z).
\end{eqnarray}
We take the two-dimensional potential to be $U(x,y)=\frac{1}{2}m^*\omega^2_x x^2+\frac{1}{2}m^*\omega^2_y y^2+
V_p(x,y)$.
The strengths of the harmonic potentials are denoted by $\omega_x$ and $\omega_y $.   They 
may be varied by {\it changing gate potentials} of the quantum dot system. 
The characteristic lengths scales along x- and y-axis are $R_{x,y}=\sqrt{\frac{\hbar}{m^*\omega_{x,y}}}$.
The potential  $V_p(x,y)$ is chosen in such a way that the parity symmetry of electron wavefunctions is broken.
We can choose it to be $V_p(x,y)=\epsilon' y$.  It  represents an electric field along the y-axis
and its strength $\epsilon'$ may be {\it varied electrically}. 

In a periodic crystal potential of a semiconductor 
the spin orbit interaction has two contributions.
The Rashba spin orbit term\cite{ras} is 
\begin{eqnarray}
H_\mathrm{R}=c_\mathrm{R} \left( \sigma_x k_y -\sigma_y k_x \right).
\end{eqnarray}
Here $\sigma_{x,y}$ are Pauli spin matrices and $k_{x,y}$ are momentum operators ($k_x=\frac{1}{i}\frac{d}{dx}$
and similarly with $k_y$.).  
The constant $c_R$ {\it depends on the external electric field} $E$ applied along the z-axis.
The Dresselhaus spin orbit term\cite{dre} is
\begin{eqnarray}
H_\mathrm{D}=c_\mathrm{D}\left( 
\left( \sigma_x k_x \left(k_y^2-k_z^2 \right) \right)
+\left( \sigma_y k_y \left(k_z^2-k_x^2 \right) \right)
\right).
\label{Dresselhaus}
\end{eqnarray}
There is another term of the form  $ \sigma_z \langle k_z\rangle \left(k_x^2-k_y^2 \right) $ in the  Dresselhaus spin orbit term
but it vanishes  since the expectation value
$\langle k_z\rangle =\langle f(z)|k_z|f(z)\rangle =0$ for the first subband  along  z-axis.
The  constant $c_D$  represents breaking of inversion symmetry by the crystal  in zinc blende structures. 
The total single electron Hamiltonian of  semiconductor quantum dot
is
$H_S=H_\mathrm{K}+H_\mathrm{R}+H_D$.
When several electrons are present in the dot
the total Hamiltonian consists of  single particle Hamiltonians and two-particle interaction terms  $H=H_0+V_{int}$, where
$H_0=\sum_iH_i$   and $V_{int}=\frac{1}{2}\sum_{i\neq j}V(\vec{r}_i-\vec{r}_j)$
with $V(\vec{r}_i-\vec{r}_j)=\frac{e^2}{\epsilon|\vec{r}_i-\vec{r}_j|}$.
($H_i$ is the single particle Hamiltonian for the ith electron and $\epsilon$
is the background dielectric constant).
The total many electron Hamiltonian is invariant under time reversal symmetry.

\section{Breaking of parity symmetry of wavefunctions}
The eigenenergies of the single electron Hamiltonian is depicted in Fig.\ref{fig:single}.
Each energy is doubly degenerate.
An  eigenstate of the Hamiltonian $H_S$ may be expanded as a linear combination of the eigenstates of $H_K$.
A lowest energy state can be written as 
\begin{eqnarray}
| \phi_1 \rangle&=&\sum_{m n}c_{m n\uparrow }|m n \uparrow \rangle 
+\sum_{m n}c_{m n\downarrow }|m n \downarrow \rangle .
\end{eqnarray}
In the basis states $|m n \sigma \rangle $ the quantum number $m (n)$ and $\sigma$ label 
the harmonic oscillator levels along  the x-axis (y-axis) and
the component of electron spin.  The subband wavefunction $f(z)$ is suppressed in the notation $|m n \sigma \rangle$.
The wavefunction $|\phi_1\rangle$  can also be written as a column vector $|\phi_1\rangle=
 \left(
  \begin{array}{c}
   F_{\uparrow} \\
    F_{\downarrow}
     \end{array}
      \right)$,
where $ F_{\uparrow}=\sum_{m n}c_{m n\uparrow }|m n  \rangle$   and 
$F_{\downarrow}=\sum_{m n}c_{m n\downarrow }|m n  \rangle$.  The expansion coefficients 
satisfy a matrix equation
\begin{eqnarray}
\sum_{m'n'\sigma'}\langle mn\sigma|H|m'n'\sigma'\rangle c_{m'n'\sigma'}=Ec_{mn\sigma}.
\end{eqnarray}
The single electron wavefunctions used in this section are  self-consistent Hartree-Fock
 single electron states if $H_S$ is replaced by the HF single electron Hamiltonian.
In the absence of the Zeeman term the single particle  Hamiltonian is invariant under time reversal symmetry:
$\vec{k}\rightarrow -\vec{k}$ and $\vec{\sigma}\rightarrow -\vec{\sigma}$.
The time reversal operator is 
$T=-i \sigma_y C$,
where the operator $C$ stands for complex conjugation. The time reversed state of $ | \phi \rangle$ is
\begin{eqnarray}
| \overline{\phi}_1 \rangle=T | \phi_1 \rangle
=-\sum_{m n}c_{m n \downarrow}^* |m n \uparrow \rangle
+\sum_{m n} c_{m n\uparrow}^* |m n \downarrow \rangle.
\end{eqnarray}
This can also be written as a column vector $|\phi_1\rangle=
\left(
\begin{array}{c}
- F_{\downarrow}^* \\
F_{\uparrow}^*
\end{array}
\right)$.
Note that $T^2| \phi \rangle=-| \phi \rangle$.
These two states are degenerate and   orthonormal.  
We have suppressed the Bloch wavefunction of the conduction band in applying the time reversal
operator since it is unaffected by the operator $T$. Our wavefunctions are 
all effective mass wavefunctions  and only the  conduction band Bloch
wavefunction at $\vec{k}=0$ is relevant.

Even in the presence of doubly degenerate states a matrix Berry phase may not always exist.
It was  shown numerically that only distorted  two-dimensional harmonic potentials
breaking the parity symmetry generate matrix Berry phases\cite{yang1}.  Here reasons for this effect for the single electron 
case is given in detail.
In this argument only one electron is assumed to be present in the dot, $c_D=0$, and $V_p(x,y)=0$.
The adiabatic parameters are $\lambda_1=c_R$ and $\lambda_2=\omega_x$.  
Let us consider $H_R$ as a perturbation and apply a  perturbation theory to $ | 00\uparrow \rangle$
\begin{eqnarray}
&&|\phi \rangle \approx |00\uparrow \rangle+ 
\sum_{mn}\frac{|mn\downarrow \rangle \langle mn\downarrow|H_R|00\uparrow \rangle}{E_{00}-E_{mn}}\nonumber\\
&+&\sum_{mn}\sum_{m'n'}\frac{|mn\uparrow \rangle \langle mn\uparrow|H_R|m'n'\downarrow \rangle}{E_{00}-E_{mn}}
\frac{ \rangle \langle m'n'\downarrow|H_R|00\uparrow \rangle}{E_{00}-E_{m'n'}}\nonumber\\
&+&....,
\label{eq:symm}
\end{eqnarray}
where $E_{nm}=(m+1/2)\hbar\omega_x+(n+1/2)\hbar\omega_y$.  The wavefunction of $|00\uparrow \rangle$ has even parity. 
The matrix element $\langle mn\downarrow|H_R|00\uparrow \rangle$ couples $|00\uparrow \rangle$  only to
odd parity states $ |mn\downarrow \rangle  $.
Therefore the wavefunction corresponding to the second term in Eq.(\ref{eq:symm}) has  odd parity with spin down. 
The wavefunction of the third term  in Eq.(\ref{eq:symm})  has even parity with spin up.
This  perturbation series
indicates
that the exact lowest energy state has the form
$|\phi_1\rangle=
\left(
\begin{array}{c}
F_{e} \\
F_{o}
\end{array}
\right)$,
where $F_{e}$ and $F_{o}$ even and odd functions.
The time reversed wavefunction is
 $|\overline{\phi}_1\rangle=
 \left(
 \begin{array}{c}
 -F_{o}^* \\
 F_{e}^*
 \end{array}
 \right)$.
For this particular pair of degenerate states the off-diagonal elements of non-Abelian vector potentials are zero:
\begin{eqnarray}
&&\langle \phi_1 |\frac{\partial}{\partial \lambda_k} |\overline{\phi}_1 \rangle
=-\langle F_{e} |\frac{\partial}{\partial \lambda_k} |F_{o}^* \rangle
+\langle F_{o} |\frac{\partial}{\partial \lambda_k} |F_{e}^* \rangle\nonumber\\
&=&-\sum_{q,p}a_k^* \frac{\partial b_p}{\partial \lambda_k}\langle \varphi_{q} |\varphi_{p} \rangle
+\sum_{q,p}b_p^* \frac{\partial a_q}{\partial \lambda_k}\langle \varphi_{p} |\varphi_{q} \rangle
= 0\nonumber\\
\end{eqnarray}
Here 
$F_e=\sum_q a_q\varphi_{q}$ with  $q=(m,n)$ such that  the two-dimensional harmonic wavefunctions  $\varphi_{q}$ are even functions.
The function  $F_o=\sum b_p\varphi_{p}$ with $p=(m',n')$ 
are such that    $\varphi_{p}$ are odd functions
(Note that one can show $\langle \varphi_q |\frac{\partial}{\partial \lambda_k} |\varphi_{p} \rangle=0$).
Since the off-diagonal non-Abelian vector potentials are zero the matrix Berry phase is absent.
In what follows a distorted two-dimensional harmonic potential is assumed.
The adiabatic parameters are $c_R$  and $\epsilon'$.

\section{Computational scheme to include many electron physics}

\begin{figure}[hbt]
\begin{center}
\includegraphics[width = 0.4 \textwidth]{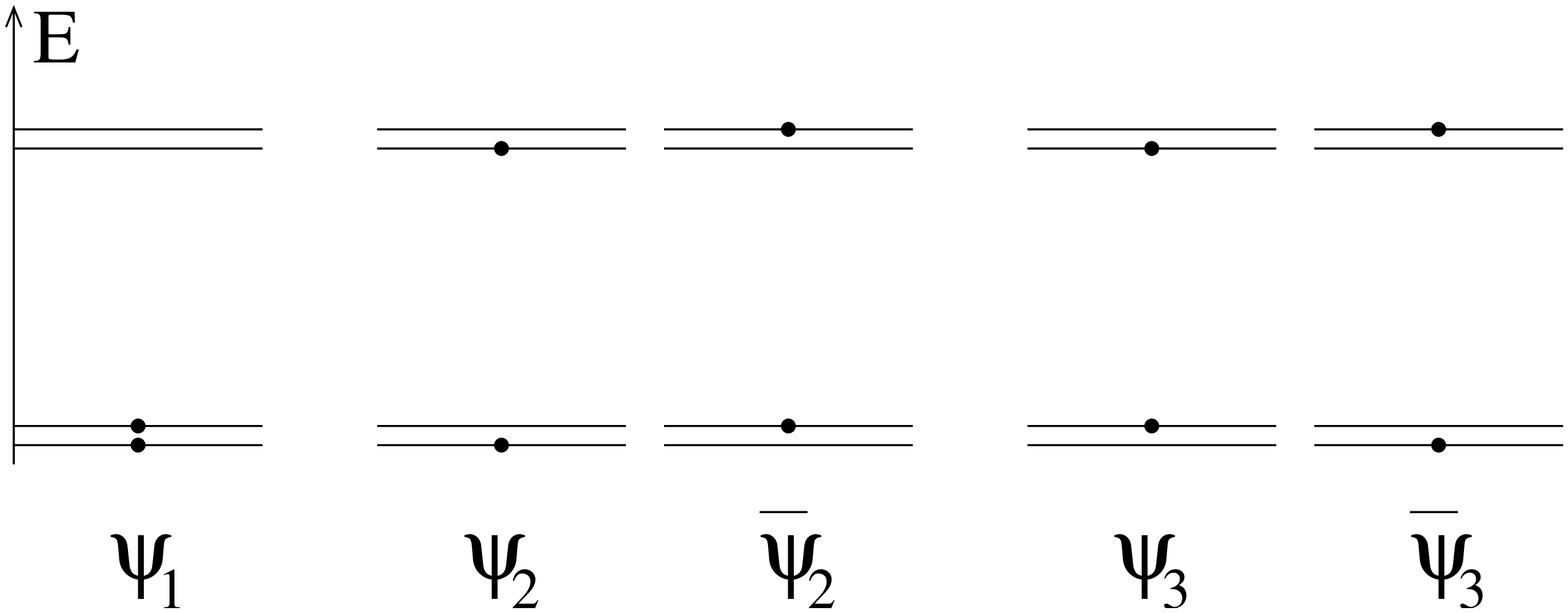}
\caption{Five many body basis states in the two electron Hilbert space. 
The basis state $\Psi_1 $ is the time reversed state of itself.
}
\label{fig:many2}
\end{center}
\end{figure}
A computational scheme that includes systematically many electron physics
is explained here.
We  find  an efficient computational scheme of computing eigenstates and eigenvalues of the many electron
Hamiltonian of a semiconductor quantum dots with an arbitrary shape of the confinement potential.
We follow  the following steps.

Single electron Hamiltonian is first solved in the presence of spin-orbit terms.
This can be done by following the steps given in Sec.III.  The number of single electron basis states $|mn\sigma\rangle$
can be truncated to a reasonable number by comparing the relative magnitude of
the single particle energy spacing and the strength of the Coulomb interaction.

Using {\it these } single electron eigenstates Slater determinant states $\Psi_i$ and their time reserved  
states $\overline{\Psi}_i$ are formed.  $\overline{\Psi}_i$ is obtained from $\Psi_i$ 
by replacing the single electron wavefunctions
with their time reversed states: $\phi_k \rightarrow \overline{\phi}_k$.  It will be  inefficient to
use two dimensional harmonic potential wavefunction $\varphi_{mn}$ to build these Slater determinant states.
The following relation hold for two arbitrary states $\Psi_a$ and $\Psi_b$:  
$\langle T\Psi_a|T\Psi_b \rangle=\langle \Psi_b|\Psi_a\rangle$.
From this  relation  it follows that   the Hamiltonian matrix elements have the following properties:
$\langle \Psi_i|H|\Psi_i\rangle=\langle \overline{\Psi}_i|H|\overline{\Psi}_i\rangle$,
$\langle \Psi_i|H|\overline{\Psi}_j\rangle=-\langle \Psi_j|H|\overline{\Psi}_i\rangle$,
and $\langle \Psi_i|H|\Psi_j\rangle=\langle \overline{\Psi}_i|H|\overline{\Psi}_j\rangle^*$.

When there are odd  number of electrons the basis vectors of   many electron
Hamiltonian are chosen to include time reversed pairs: 
$\{\Psi_1,\overline{\Psi}_1,\Psi_2,\overline{\Psi}_2,...   \}$, see Fig. \ref{fig:many}.
The eigenstates are doubly degenerate
in the odd electron case.
Let us show this for the three electron case
using  the following Slater determinant wavefunctions as the basis vectors in the many body Hilbert space
$\{\Psi_1,\overline{\Psi}_1,\Psi_2,\overline{\Psi}_2   \}$.   These states are depicted in Fig.\ref{fig:many}.
The wavefunctions of the lowest energy degenerate pair are
$\Phi=a\Psi_1+b\overline{\Psi}_1+c\Psi_2+d\overline{\Psi}_2$ and 
$\overline{\Phi}=T\Phi=a^*\overline{\Psi}_1-b^*\Psi_1+c^*\overline{\Psi}_2-d^*\Psi_2$.
Clearly $\langle\Phi |H|\Phi\rangle=\langle\overline{\Phi} |H|\overline{\Phi}\rangle$.
Note $\Phi$ and $\overline{\Phi}$ are a time reversed pair,  orthogonal, and satisfies $T^2\Phi=-\Phi$.
The two eigenvalues of $H$ are each doubly degenerate.

When there are even   number of electrons the basis vectors of the   many electron
Hamiltonian are chosen as $\{\Psi_1,\Psi_2,\overline{\Psi}_2,\Psi_3,\overline{\Psi}_3,...   \}$.
In this scheme  the groundstate is non degenerate.
This can be shown shown, for example, in an  approximation that includes five many body basis vectors
in the presence of two electrons, see Fig. \ref{fig:many2}.
The groundstate can then be written as a linear combination
$\Phi=a\Psi_1+b\Psi_2+c\overline{\Psi}_2+d\Psi_3+e\overline{\Psi}_3$.  Its time reversed state is
$\overline{\Phi}=a^*\Psi_1+b^*\overline{\Psi}_2-c^*\Psi_2+d^*\overline{\Psi}_3-e^*\Psi_3$.
Clearly these two states are {\it not } orthogonal.
Since $\langle \overline{\Phi} |H|\overline{\Phi} \rangle=\langle T^2\Phi |H|T^2\Phi \rangle$
and $T^2  |\Phi \rangle  \neq -|\Phi \rangle$ one finds 
that the energy expectation values are non degenerate,
$\langle \overline{\Phi} |H|\overline{\Phi} \rangle\neq \langle \Phi |H|\Phi \rangle$, i.e., 
time reversed states are non degenerate.
The matrix Berry phase is thus absent.
In the simplest possible approximation the  groundstate $\Psi$ of two electron system 
is  given by the single Slater determinant made of 
the  single particles states  $\{\phi_1,\overline{\phi_1} \}$.
In this case  the initial state should return to itself after the 
adiabatic cycle.  It is instructive to verify this explicitly by transforming each single particles states.
The  initial state  is 
\begin{eqnarray}
|\Psi(0)\rangle=\phi_1(\vec{r}_1)\overline{\phi}_1(\vec{r}_2)-\phi_1(\vec{r}_2)\overline{\phi}_1(\vec{r}_1).
\end{eqnarray}
After a cycle the single electron wavefunctions transform into
\begin{eqnarray}
\phi_1 \rightarrow a_1\phi_1+b_1\overline{\phi}_1  \textrm{ and  }  \overline{\phi}_1\rightarrow a_2\phi_1+b_2\overline{\phi}_1.
\end{eqnarray}
The transformed two electron Slater determinant is
$|\Psi(T)\rangle=(a_1b_2-b_1a_2)(\phi_1(\vec{r}_1)\overline{\phi}_1(\vec{r}_2)-\phi_1(\vec{r}_2)\overline{\phi}_1(\vec{r}_1))$,
which is
$ (a_1b_2-b_1a_2)|\Psi(0)\rangle$.
Since the transformed single electron states are time reversed 
states of each other $(a_2,b_2)=(-b_1^*,a_1^*)$.  So the factor in the transformed wavefunction is
$a_1b_2-b_1a_2=|a_1|^2+|b_1|^2=1$.

\section{Simplest possible approximation with antisymmetry of wavefunctions }

Now consider odd number of electrons, for example, three electrons in a dot.
In the simplest possible approximation with fermionic antisymmetry
two many body basis states, $\Psi_1$ and $\overline{\Psi}_1$,  can be included.
These are Slater determinant states  and  are depicted in Fig.\ref{fig:many}.
This approximation is well justified when the typical single electron energy spacing is 
larger than the characteristic Coulomb energy.
Clearly $\Psi_1$ and $\overline{\Psi}_1$ are orthogonal $\langle \Psi_1|\overline{\Psi}_1  \rangle=0$.
The  Hamiltonian matrix is
\begin{eqnarray}
H
&=&
\left(
\begin{array}{cc}
\langle \Psi_1|H|\Psi_1\rangle & \langle \Psi_1|H|\overline{\Psi}_1 \rangle\\
 \langle \overline{\Psi}_1|H|\Psi_1\rangle &   \langle \overline{\Psi}_1|H|\overline{\Psi}_1\rangle
\end{array}
\right). \nonumber \\
\end{eqnarray}
The off-diagonal matrix elements of the Hamiltonian matrix  
is zero
since $\langle \Psi_1|H|\overline{\Psi}_1 \rangle=-\langle \Psi_1|H|\overline{\Psi}_1 \rangle$,
which follows from 
 $\langle T\Psi_a|T\Psi_b \rangle=\langle \Psi_b|\Psi_a\rangle$.
It is instructive to show this  explicitly by evaluating the matrix element
\begin{eqnarray}
&&\langle\Psi_1 |V_{int}|\overline{\Psi}_1 \rangle \nonumber\\
&=&[\langle 23 |V|24 \rangle-\langle 23 |V|42 \rangle]+[\langle 13 |V|14 \rangle-\langle 13 |V|4 1 \rangle].\nonumber\\
\end{eqnarray}
Here the first and third terms are direct Coulomb matrix elements while 
the second and fourth terms are their  exchange counter parts.  The direct Coulomb matrix elements are zero
since $\phi_3^*(\vec{r})\phi_4(\vec{r})=\phi_3^*(\vec{r})\overline{\phi}_3(\vec{r})=0$.
So only their exchange counter parts remain, which are given by  
\begin{eqnarray}
\langle 23 |V|42 \rangle=\int dr_1dr_2 \overline{\phi}_1^*(\vec{r}_1)\phi_3^*(\vec{r}_2)V(\vec{r}_1-\vec{r}_2)
\overline{\phi}_3(\vec{r}_1) \overline{\phi}_1(\vec{r}_2).\nonumber\\
\end{eqnarray}
and 
\begin{eqnarray}
\langle 13 |V|41 \rangle=\int dr_1dr_2 \phi_1^*(\vec{r}_1)\phi_3^*(\vec{r}_2)V(\vec{r}_1-\vec{r}_2)
\overline{\phi}_3(\vec{r}_1) \phi_1(\vec{r}_2).\nonumber\\
\end{eqnarray}
One can show 
\begin{eqnarray}
\langle\Psi_1 |V_{int}|\overline{\Psi}_1 \rangle 
=-\langle 23 |V|42 \rangle-\langle 13 |V|4 1 \rangle=0.
\end{eqnarray}
This is because 
\begin{eqnarray}
\langle 23 |V|42 \rangle=-\langle 13 |V|4 1 \rangle,
\end{eqnarray}
which follows from  $\phi_2^*=\overline{\phi}_1^*$, $\phi_4=\overline{\phi}_3$, and 
\begin{eqnarray}
\overline{\phi}_1^*(\vec{r})\overline{\phi}_3(\vec{r})&=&
F_{1\downarrow}(\vec{r})F_{3\downarrow}^*(\vec{r})+F_{1\uparrow}(\vec{r})F_{3\uparrow}^*(\vec{r}),\nonumber\\
\phi_3^*(\vec{r})\overline{\phi}_1(\vec{r})&=&
-F_{3\uparrow}^*(\vec{r})F_{1\downarrow}^*(\vec{r})+F_{3\downarrow}^*(\vec{r})F_{1\uparrow}^*(\vec{r}),\nonumber\\
\phi_1^*(\vec{r})\overline{\phi}_3(\vec{r})&=&
-F_{1\uparrow}^*(\vec{r})F_{3\downarrow}^*(\vec{r})+F_{1\downarrow}^*(\vec{r})F_{3\uparrow}^*(\vec{r}),\nonumber\\
\phi_3^*(\vec{r})\phi_1(\vec{r})&=&
F_{3\uparrow}^*(\vec{r})F_{1\uparrow}(\vec{r})+F_{3\downarrow}^*(\vec{r})F_{1\downarrow}(\vec{r}).
\end{eqnarray}
Note that the second and third terms are equal except for the minus sign.
The off-diagonal elements are thus zero, and 
\begin{eqnarray}
H
&=&
\left(
\begin{array}{cc}
\langle \Psi_1|H|\Psi_1\rangle & 0\\
 0 &   \langle \overline{\Psi}_1|H|\overline{\Psi}_1\rangle
\end{array}
\right). \nonumber \\
\end{eqnarray}
This implies that  $\Psi_1$ and $\overline{\Psi}_1$ are eigenstates and are degenerate in energy.

During the adiabatic change the wavefunction of three electrons evolves as 
$|\Psi'(t)\rangle =c_1(t)|\Psi_1(t)\rangle +c_2(t)|\overline{\Psi}_1(t)\rangle$. 
The coefficients $c_1(t)$  and  $c_2(t)$ satisfy the Schroedinger equation  Eq.(\ref{eq:time_Schrod}) governed by 
the non-Abelian vector potentials
\begin{eqnarray}
(A_k)_{1,1}=i\langle \Psi_1|\frac{\partial}{\partial\lambda_k}|\Psi_1  \rangle,\nonumber\\
(A_k)_{1,2}=i\langle \Psi_1|\frac{\partial}{\partial\lambda_k}|\overline{\Psi}_1  \rangle,\nonumber\\
(A_k)_{2,1}=i\langle \overline{\Psi}_1|\frac{\partial}{\partial\lambda_k}|\Psi_1  \rangle, \nonumber\\
(A_k)_{2,2}=i\langle \overline{\Psi}_1|\frac{\partial}{\partial\lambda_k}|\overline{\Psi}_1  \rangle.
\end{eqnarray}
It can be shown using Eqs.(\ref{eq:diag}) and (\ref{eq:matrixover})
that these can be written in terms of  the non Abelian vector potentials for single particle states
\begin{eqnarray}
(A_k)_{1,1}&=&(a_k)_{1,1}+(a_k)_{2,2}+(a_k)_{3,3},\nonumber\\
(A_k)_{1,2}&=&(a_k)_{3,4},\nonumber\\
(A_k)_{2,1}&=&(a_k)_{4,3},\nonumber\\
(A_k)_{2,2}&=&(a_k)_{1,1}+(a_k)_{2,2}+(a_k)_{4,4},
\label{eq:nonabel}
\end{eqnarray}
where 
the non-Abelian vector potentials involving the lowest energy single electron states are
\begin{eqnarray}
(a_k)_{1,1}=i\langle \phi_1|\frac{\partial}{\partial\lambda_k}|\phi_1  \rangle,\nonumber\\
(a_k)_{2,2}=i\langle \overline{\phi}_1|\frac{\partial}{\partial\lambda_k}|\overline{\phi}_1  \rangle,
\end{eqnarray}
and involving the second lowest single electron states are
\begin{eqnarray}
(a_k)_{3,3}=i\langle \phi_2|\frac{\partial}{\partial\lambda_k}|\phi_2  \rangle,\nonumber\\
(a_k)_{4,4}=i\langle \overline{\phi}_2|\frac{\partial}{\partial\lambda_k}|\overline{\phi}_2  \rangle,\nonumber\\
(a_k)_{3,4}=i\langle \phi_2|\frac{\partial}{\partial\lambda_k}|\overline{\phi}_2  \rangle,\nonumber\\
(a_k)_{4,3}=i\langle \overline{\phi}_2|\frac{\partial}{\partial\lambda_k}|\phi_2  \rangle.
\end{eqnarray}

The solution of the time dependent Schroedinger equation has the form
$\left(
\begin{array}{c}
c_{1}(t) \\
c_{2}(t)
\end{array}
\right)=
A(t)\left(
\begin{array}{c}
B_{1}(t) \\
B_{2}(t)
\end{array}
\right)$.
The function $A(t)$  depends on the electrons in the {\it first} shell.  It 
satisfies $f(t)A(t)=-i\hbar\frac{dA}{dt}$,
where
$f(t)=\sum_k ((a_k)_{1,1}+(a_k)_{2,2})\frac{d\lambda_k}{dt}$.
The solution of this differential equation is a phase factor
\begin{eqnarray}
A(t)=e^{\frac{i}{\hbar} f(t)}.
\end{eqnarray}
Note that 
$(a_k)_{1,1}+(a_k)_{2,2}$ appears in the diagonal 
elements of the non Abelian vector potentials $(A_k)_{1,1}$  and  $(A_k)_{2,2}$.
The functions $B_{1}(t)$ and $B_{2}(t)$ 
are the solutions of the time dependent Schroedinger equation
\begin{eqnarray}
i \hbar \dot{B}_i=-\sum_j \overline{A}_{i j} B_j \qquad i=1,2.
\end{eqnarray}
with the following non-Abelian gauge potentials.
\begin{eqnarray}
(\overline{A}_k)_{1,1}&=&(a_k)_{3,3},\nonumber\\
(\overline{A}_k)_{1,2}&=&(a_k)_{3,4},\nonumber\\
(\overline{A}_k)_{2,1}&=&(a_k)_{4,3},\nonumber\\
(\overline{A}_k)_{2,2}&=&(a_k)_{4,4}.
\end{eqnarray}
This is precisely the  non Abelian vector potentials for the {\it second} energy shell.
The probability to find an electron in the state
$|\phi_3\rangle (|\phi_4\rangle)$ is $|c_1(T)|^2=|B_1(T)|^2 (|c_2(T)|^2=|B_2(T)|^2 )$.
So these probabilities 
are  {\it independent} of   the non-Abelian vector potentials of the first energy shell.
In addition it should be noted that 
the antisymmetry of the many electron wavefunction does {\it not} 
change the value of the matrix Berry phase obtained from the wavefunctions without the antisymmetry, see Eq.(\ref{eq:berry}).
This is the  main conclusion of this section.

\section{Correlation effects }

Correlation
effects can be taken into account by writing  many body eigenstates as a linear combination
of  single Slater determinant wavefunctions 
$\{\Phi_1,\Phi_2,\Phi_3,\Phi_4,...\}=\{\Psi_1,\overline{\Psi}_1,\Psi_2,\overline{\Psi}_2,...\}$. These 
wavefunctions arranged in the increasing order of the total kinetic energy $E_1< E_2< ...$.
Note that $ \langle \Phi_i|\Phi_j \rangle=\delta_{i,j}$ for all i and j. 
If the groundstate is doubly degenerate
one of the degenerate groundstates can be written as 
\begin{eqnarray}
|\Phi\rangle=\sum_{i=1}^Mc_i|\Phi_i\rangle,
\end{eqnarray}
with
\begin{eqnarray}
\sum_{j=1}^M  \langle \Phi_i|H|\Phi_j \rangle c_j=Ec_i.
\end{eqnarray}
Then its time reversed state
has the form 
\begin{eqnarray}
|\overline{\Phi}\rangle=T\Phi=\sum_{i=1}^{M}c_i^*|\overline{\Phi}_i\rangle
\end{eqnarray}
since $T$ is an antiunitary operator.  
Note that $\overline{\Phi}_i$ are  ordered as follows
$\{\overline{\Phi}_1,\overline{\Phi}_2,\overline{\Phi}_3,\overline{\Phi}_4,...\}=
\{\overline{\Psi}_1,\Psi_1,\overline{\Psi}_2,\Psi_2,...\}$.
The states $|\Phi\rangle$ and $|\overline{\Phi}\rangle$  are eigenstates of the truncated Hamiltonian with the dimension $M$.
(Note that as $M\rightarrow \infty$ the energy $E$ will approach the exact eigenvalue.).
Some of the Slater determinant states $|\Psi_i\rangle$ are depicted in Fig.\ref{fig:many}.
Clearly $|\Phi\rangle$ and $|\overline{\Phi}\rangle$ are orthogonal. The matrix element between them 
is zero: $\langle \Phi|H|\overline{\Phi} \rangle=E\langle \Phi|\overline{\Phi} \rangle$=0.
These two states are degenerate since  $\langle\Phi |H|\Phi  \rangle=\langle \overline{\Phi}|H|\overline{\Phi}  \rangle$.

The non-Abelian vector potentials are
\begin{eqnarray}
(A_k)_{1,1}=i\langle \Phi|\frac{\partial}{\partial\lambda_k}|\Phi  \rangle,\nonumber\\
(A_k)_{1,2}=i\langle \Phi|\frac{\partial}{\partial\lambda_k}|\overline{\Phi}  \rangle,\nonumber\\
(A_k)_{2,1}=i\langle \overline{\Phi}|\frac{\partial}{\partial\lambda_k}|\Phi  \rangle, \nonumber\\
(A_k)_{2,2}=i\langle \overline{\Phi}|\frac{\partial}{\partial\lambda_k}|\overline{\Phi}  \rangle.
\label{eq:nonab}
\end{eqnarray}
In terms of many body basis states they can be written as
\begin{eqnarray}
(A_k)_{1,1} &=&i\sum_ic_i^*\frac{\partial{c_i}}{\partial\lambda_k}
+i\sum_{i,j}c_i^*c_j\langle \Phi_i|\frac{\partial}{\partial\lambda_k}|\Phi_j  \rangle,\nonumber\\
(A_k)_{2,2} &=&i\sum_ic_i\frac{\partial{c_i^*}}{\partial\lambda_k} +i\sum_{i,j}c_ic_j^*
\langle \overline{\Phi}_i|\frac{\partial}{\partial\lambda_k}|\overline{\Phi}_j  \rangle, \nonumber\\
(A_k)_{1,2}& =&i\sum_{i,j}c_i^*c_j^*\langle \Phi_i|\frac{\partial}{\partial\lambda_k}|\overline{\Phi}_j  \rangle,\nonumber\\
(A_k)_{2,1} &=&i\sum_{i,j}c_ic_j\langle \overline{\Phi}_i|\frac{\partial}{\partial\lambda_k}|\Phi_j  \rangle.
\end{eqnarray}
In the three electron case  with the basis vectors given in Fig.\ref{fig:many} 
the diagonal matrix elements  
$i\langle \Phi_i|\frac{\partial}{\partial\lambda_k}|\Phi_i  \rangle$ and 
$i\langle \overline{\Phi}_i|\frac{\partial}{\partial\lambda_k}|\overline{\Phi}_i  \rangle$ are
given by 
\begin{eqnarray}
i\langle \Psi_{1}|\frac{\partial}{\partial\lambda_k}|\Psi_{1}  \rangle&=&(a_k)_{1,1}+(a_k)_{2,2}+(a_k)_{3,3},\nonumber\\
i\langle \overline{\Psi}_{1}|\frac{\partial}{\partial\lambda_k}|\overline{\Psi}_{1}  \rangle
&=&(a_k)_{1,1}+(a_k)_{2,2}+(a_k)_{4,4},\nonumber\\
i\langle \Psi_{2}|\frac{\partial}{\partial\lambda_k}|\Psi_{2}  \rangle&=&(a_k)_{1,1}+(a_k)_{2,2}+(a_k)_{5,5},\nonumber\\
i\langle \overline{\Psi}_{2}|\frac{\partial}{\partial\lambda_k}|\overline{\Psi}_{2}  \rangle&=&(a_k)_{1,1}+(a_k)_{2,2}+(a_k)_{6,6}.
\label{eq:nonabel}
\end{eqnarray}
Similarly one can find off-diagonal matrix elements  
\begin{eqnarray}
i\langle \Psi_{1}|\frac{\partial}{\partial\lambda_k}|\overline{\Psi}_{1}  \rangle&=&(a_k)_{3,4},\nonumber\\
i\langle \Psi_{2}|\frac{\partial}{\partial\lambda_k}|\overline{\Psi}_{2}  \rangle&=&(a_k)_{5,6},\nonumber\\
i\langle \Psi_{1}|\frac{\partial}{\partial\lambda_k}|\Psi_{2}  \rangle&=&(a_k)_{3,5},\nonumber\\
i\langle \Psi_{1}|\frac{\partial}{\partial\lambda_k}|\overline{\Psi}_{2}  \rangle&=&(a_k)_{3,6},\nonumber\\
i\langle \overline{\Psi}_{1}|\frac{\partial}{\partial\lambda_k}|\Psi_{2}  \rangle&=&(a_k)_{4,5},\nonumber\\
i\langle \overline{\Psi}_{1}|\frac{\partial}{\partial\lambda_k}|\overline{\Psi}_{2}  \rangle&=&(a_k)_{4,6}.\nonumber\\
\label{eq:result}
\end{eqnarray}
These results follow from Eq.(\ref{eq:matrixover}).
Usually the  non Abelian
vector potentials are defined between two degenerate states, see Eq.(\ref{eq:vector_pot}).
When electron correlations are important the  matrix elements between two single electron states belonging to
{\it different } energy shells 
are relevant (see the last four equations in Eq.(\ref{eq:result})).
We will call these   inter shell non-Abelian
vector potentials. 
Physically the mixing  between the second and third shells is facilitated by them.
Note that $(a_k)_{j,i}=(a_k)_{i,j}^*$.

\section{Discussions and conclusions}

We have developed  a  scheme of computing eigenstates and eigenvalues of the  many electron
Hamiltonian of a semiconductor quantum dot with an arbitrary shape of the confinement potential.
The basis vectors of this  many electron
Hamiltonian are chosen to include time reversed pairs of Slater determinant states. 
In this scheme  the groundstate is non degenerate
when  an even number of electrons are present  while 
the eigenstates are doubly degenerate when 
an odd number of electrons are present.

Using this computational scheme we have investigated matrix Berry phases 
when several electrons are present in a semiconductor quantum dot with
spin-orbit coupling terms.  
When an odd number of electrons is present a matrix Berry phase exists.   
The effect of the antisymmetry of
many electron wavefunctions does not affect the value of  the  matrix Berry phase.
Electron correlation effects can be treated by including  the inter shell coupling.
In this case  the many body non-Abelian vector potentials  can be written as a sum of those of the single electron
non-Abelian vector potentials,
provided that  the  inter shell non-Abelian
vector potentials, $i\langle \phi_p|\frac{\partial}{\partial\lambda}|\phi_q \rangle$ with $\phi_p$ and $\phi_q $ non degenerate, 
are introduced to represent the shell coupling. It would be interesting to investigate the role of these
inter shell non-Abelian
vector potentials in parabolic dots.
Optical dipole transitions\cite{hei,dip} can be used to detect matrix Berry phases even when several electrons interact
strongly with each other.
This is because  the dipole oscillator strength will depend on the matrix Berry phase\cite{yang1}. 
However, our results in Sec. V
indicate the  matrix Berry phases will be independent of many body physics when the  energy level 
spacing between the shells is larger than the characteristic Coulomb
energy.  
In  self assembled quantum dots with wetting layers this condition is satisfied\cite{pet}.
In this case only the intra shell non-Abelian vector potentials, 
$i\langle \phi_p|\frac{\partial}{\partial\lambda}|\phi_q \rangle$ with $\phi_p$ and $\phi_q $ 
degenerate, are relevant, and the value of the matrix Berry phase can be  calculated as in ref.\cite{yang1}.
We leave it as a future study to investigate  numerical values of  the matrix Berry phases
for strongly interacting electrons.

In II-VI semiconductors the Rashba term is expected to be larger than the
the Dresselhaus coupling and  in III-V semiconductors the
opposite is true\cite{ras}.  
The typical value of  the energy scale associated with
the Rashba constant  depends on the electric field applied along the z-axis
and the semiconductor material: it is of  order  $E_R=c_R/ R\sim 0.01-10meV$, where the   length
scale $R\sim 100 \AA$ is the lateral dimension of the quantum dot.
The second adiabatic parameter $E_p$ represents the strength of the distortion
potential $\epsilon' y$: the expectation value of the distortion potential is $E_p=\langle 0|\epsilon'
y|1\rangle$.  Its magnitude is of order $1-10meV$, depending on the electric field applied along
the y-axis.

In self-assembled InAs quantum dots\cite{pet} the level spacing can be  larger than the characteristic Coulomb energy, 
and our investigation  shows that in such a case the many problem essentially reduces to a single particle
problem (The energy $E_0$ associated with the length scale $R$ is
typically of order 20-40meV in self-assembled quantum dots). In this regime, as we explained above, 
the exchange effect disappears in matrix Berry phases and 
it is only the last singly occupied
electron that matters.  The presence of the other electrons 
in the inert filled shells
will  just modify
the confinement potential.  Thus
the matrix Berry phase can be   calculated
following  the method outlined in Ref[11], and below we 
give some typical values of the matrix Berry phases.  We take  an elliptic adiabatic path   given by 
$(E_R(t),E_p(t))= 
(E_{R,c}+\Delta E_R \cos(\omega t),E_{p,c}+\Delta E_p \sin(\omega t))$.
For the parameters $E_{R,c}/E_0= 2$, $E_{p,c}/E_0=1$,  $\Delta E_p/E_0= 0.9 $, and $\hbar\omega/E_0=0.1 $
we find 
$(|c_1(T)|^2,|c_2(T)|^2)=(0.7971,0.2036), \ (0.9138,0.0867),\ (0.9810,0.0191)$, respectively, 
for different values of   the semiaxis of the ellipse
$\Delta E_R/E_0= 1.9,\ 1.5,\ 1$ (The initial state is  $(|c_1(0)|^2,|c_2(0)|^2)=(1,0)$).
Note that the parameter for the semiaxis $\Delta E_R/E_0$ 
can be varied experimentally either by changing the Rashba field $\Delta E_R$ or the size of the dot through $E_0$.
We have verified that two  different values of the frequency  $\omega$ give the same  values of  
$(|c_1(T)|^2,|c_2(T)|^2)$ as long as the other parameters
are unchanged, confirming that the matrix Berry phase is a geometric phase\cite{Wil}.
This should be tested in any experimental investigation of matrix Berry phases.    
Experimental data of single quantum dots would be  less complicated than data of an ensemble of dots with various sizes of dots,
and can thus
be compared directly with our theoretical results.

\appendix
\section{Mathematical note}

Let $|\Psi_1\rangle$ consists of $\{\phi_1,\phi_2,\phi_3\}$ and let
$|\Psi_2\rangle$ consists of $\{\phi_1,\phi_2,\phi_4\}$:
\begin{widetext}
\begin{eqnarray}
|\Psi_1\rangle=\frac{1}{\sqrt{3!}}&\{&  \phi_1(r_1)\phi_2(r_2)\phi_3(r_3)
+\phi_3(r_1)\phi_1(r_2)\phi_2(r_3)
+\phi_2(r_1)\phi_3(r_2)\phi_1(r_3)\nonumber\\
&-&\phi_3(r_1)\phi_2(r_2)\phi_1(r_3)
-\phi_1(r_1)\phi_3(r_2)\phi_2(r_3)
-\phi_2(r_1)\phi_1(r_2)\phi_3(r_3) \}  
\end{eqnarray}
and 
\begin{eqnarray}
|\Psi_2\rangle=\frac{1}{\sqrt{3!}}&\{&\phi_1(r_1)\phi_2(r_2)\phi_4(r_3)
+\phi_4(r_1)\phi_1(r_2)\phi_2(r_3)
+\phi_2(r_1)\phi_4(r_2)\phi_1(r_3)\nonumber\\
&-&\phi_4(r_1)\phi_2(r_2)\phi_1(r_3)
-\phi_1(r_1)\phi_4(r_2)\phi_2(r_3)
-\phi_2(r_1)\phi_1(r_2)\phi_4(r_3)     \}. 
\end{eqnarray}
\end{widetext}
The diagonal elements of the non-Abelian gauge potential between the Slater determinant wavefunctions  are
\begin{eqnarray}
i\langle \Psi_{i}|\frac{\partial}{\partial\lambda_k}|\Psi_{i}  \rangle=
i\sum_p \langle \phi_{p}(\vec{r})|\frac{\partial}{\partial \lambda_k}  |\phi_{p}(\vec{r}) \rangle.
\label{eq:diag}
\end{eqnarray}
Let us find the off-diagonal non-Abelian vector potential elements  
$i\langle \Psi_1|\frac{\partial}{\partial\lambda_k}|\Psi_2  \rangle$
between the two  Slater determinant wavefunctions $| \Psi_1\rangle$ and $|\Psi_2  \rangle$.
When  $\phi_3$ is replaced with $\phi_4$ the state $|\Psi_1\rangle$ transforms into $|\Psi_2\rangle$.
We find $\langle \Psi_1|\frac{\partial}{\partial\lambda_k}|\Psi_2  \rangle=
\langle \phi_3(\vec{r})|\frac{\partial}{\partial \lambda_k}  |\phi_4(\vec{r}) \rangle$.
This result may be easily generalized to 
$N$ particle Slater determinant wavefunctions.
In general the non-Abelian vector potential element between two {\it different} Slater determinant wavefunctions,
 $|\Psi_i\rangle$ and $|\Psi_j\rangle$, is non zero when
 $|\Psi_i\rangle$ can be transformed into  $|\Psi_j\rangle$ 
by replacing
one single electron wavefunction in  $|\Psi_i\rangle$ with another  wavefunction in  $|\Psi_j\rangle$.
If these single electron wavefunctions are denoted by
$\phi_{\ell}$ with $\phi_{p}$ then 
we find 
\begin{eqnarray}
\langle \Psi_{i}|\frac{\partial}{\partial\lambda_k}|\Psi_{j}  \rangle=
\langle \phi_{\ell}(\vec{r})|\frac{\partial}{\partial \lambda_k}  |\phi_{p}(\vec{r}) \rangle.
\label{eq:matrixover}
\end{eqnarray}
If more than two single electron wavefunctions in  $|\Psi_i\rangle$ must be exchanged  with those in 
$|\Psi_j\rangle$ to transform $|\Psi_i\rangle$
into  $|\Psi_j\rangle$
then  $\langle \Psi_{i}|\frac{\partial}{\partial\lambda_k}|\Psi_{j}  \rangle=0$.

\begin{acknowledgments}
This work was  supported by grant No. R01-2005-000-10352-0 from the Basic Research Program
of the Korea Science and Engineering
Foundation and by Quantum Functional Semiconductor Research Center (QSRC) at Dongguk University
of the Korea Science and Engineering
Foundation. In addition this work was supported by The Second Brain 21 Project. 
\end{acknowledgments}


\begin{references}
\bibitem{Aws}D. D. Awschalom, D. Loss, and N. Samarth, Semiconductor Spintronics and Quantum Computation (Springer, Berlin, 2002).
\bibitem{ras0}E. I. Rashba and Al. L. Efros, Phys. Rev. Lett. {\bf 91}, 126405 (2003).
\bibitem{Sha}{\it Geometric Phases in Physics}, edited by A. Shapere and F. Wilczek
(World Scientific, Singapore, 1989). 
\bibitem{Wil}F. Wilczek and A Zee, Phys. Rev. Lett. {\bf 52}, 2111 (1984).
\bibitem{Zan}P. Zanardi and M. Rasetti, Phys. Lett. {\bf 264}, 94 (1999).
\bibitem{Pach}J. Pachos, P. Zanardi and M. Rasetti, Phys. Rev. A {\bf 61}, 010305(R) (2000); 
J. Pachos and  P. Zanardi, Int. J. Mod. Phys. B {\bf 15}, 1257 (2001).
\bibitem{Bern}B. A. Bernevig and S.-C. Zhang, Phys. Rev. B {\bf 71}, 035303 (2005).
\bibitem{Sere2}Yu. A. Serebrennikov, Phys. Rev. B {\bf 70}, 064422 (2004). 
\bibitem{Sol2}P. Solinas, P. Zanardi, N. Zanghi, and F. Rossi, Phys. Rev.A {\bf 67}, 062315 (2003).
\bibitem{par}D. Parodi, M.Sassetti, P. Solinas, P. Zanardi, and N. Zanghi, Phys. Rev.A {\bf 73}, 052304 (2006).
\bibitem{yang1}S.-R. Eric Yang and N.Y. Hwang,  Phys. Rev. B  {\bf 73}, 125330 (2006).
\bibitem{val}M. Valin-Rodriquez, A. Puente, and L. Serra, Phys. Rev. B   {\bf 69}, 085306  (2004).
\bibitem{yang2}S.-R. Eric Yang,  Phys. Rev. B   {\bf 74}, 075315 (2006).
\bibitem{kou}L. P. Kouvenhoven, C. M. Marcus, P. L. McEuen, S. Tarucha, R. M. Westervelt, and N. S. Wingreen, in Nato ASI
conference Proceedings, Section 5.1, edited by L. P. Kouvenhoven, G. Sch\"{o}n, and L. L. Sohn (Kluwer, Dordrecht, 1977).
\bibitem{lee}S. D. Lee, S. J. Kim, Y. B. Cho, J. B. Choi, Sooa Park, S.-R. Eric Yang, 
S. J. Lee and T. H. Zyung, Appl. Phys. Lett. 89, 023111 (2006).
\bibitem{ras}E.I. Rashba, Physica E,  {\bf 34}, 31 (2006).
\bibitem{dre}G. Dresselhaus, Phys. Rev. {\bf 100}, 580 (1955).
\bibitem{hei}D. Heitmann and J. P. Kotthaus, Phys. Today  {\bf 56} (6), 56 (1993);
\bibitem{dip}The effect of many electron physics on dipole absorption spectra are investigated in parabolic dots in the presence of spin orbit terms
in the following papers: P. Lucignano, B. Jouault, A. Tagliacozzo, and B. L. Altshuler, Phys. Rev. B   {\bf 71}, 121310(R)  (2005);
P. Pietil\"{a}inen and T. Chakraborty,  Phys. Rev. B   {\bf 73}, 155315  (2006). To detect matrix Berry
phases it is desirable  to investigate similar effects in non-parabolic dots with parity breaking terms.
\bibitem{pet}P. M. Petroff, A. Lorke, and A. Imamoglu, Phys. Today  {\bf 54} (5), 46 (2001).

\end{references}
\end{document}